\documentclass[numberedappendix]{emulateapj}
\usepackage{epsfig, natbib}

\tightenlines

\countdef\decade=200
\advance\decade by \year
\countdef\hours=201	
\advance\hours by \time
\divide\hours by 60
\countdef\mins=202
\advance\mins by \hours
\multiply\mins by 60
\multiply\hours by 100
\countdef\miltime=203
\advance\miltime by \hours
\advance\miltime by \time
\advance\miltime by -\mins

\newcommand{\msun}{M_{\odot}}
\newcommand{\hi}{H~\textsc{i}}
\newcommand{\htwo}{H$_2$}
\newcommand{\fhtwo}{f_{\rm H_2}}
\newcommand{\fhtwosim}{f_{\rm H_2,sim}}
\newcommand{\fhtwoanalyt}{f_{\rm H_2,analyt}}
\newcommand{\dfhtwo}{\Delta f_{\rm H_2}}
\newcommand{\calr}{\mathcal{R}}
\newcommand{\nhzero}{n_{\rm H,0}}
\newcommand{\nh}{n_{\rm H}}

\def\Msun{\, M_{\odot}}

\begin{document}

\title{A Comparison of Methods for Determining the Molecular Content of Model Galaxies}


\shorttitle{H$_2$ in Simulations}
\shortauthors{Krumholz \& Gnedin}

\author{Mark R.\ Krumholz\altaffilmark{1} and Nickolay Y.\ Gnedin\altaffilmark{2,3,4}}
\altaffiltext{1}{Department of Astronomy and Astrophysics, University of California, Santa Cruz, CA 95064, USA; krumholz@ucolick.org}
\altaffiltext{2}{Particle Astrophysics Center, Fermi National Accelerator Laboratory, Batavia, IL 60510, USA; gnedin@fnal.gov}
\altaffiltext{3}{Kavli Institute for Cosmological Physics, The University of Chicago, Chicago, IL 60637 USA} 
\altaffiltext{4}{Department of Astronomy and Astrophysics, The University of Chicago, Chicago, IL 60637 USA} 

\begin{abstract}
Recent observations indicate that star formation occurs only in the molecular phase of a galaxy's interstellar medium. A realistic treatment of star formation in simulations and analytic models of galaxies therefore requires that one determine where the transition from the atomic to molecular gas occurs. In this paper we compare two methods for making this determination in cosmological simulations where the internal structures of molecular clouds are unresolved: a complex time-dependent chemistry network coupled to a radiative transfer calculation of the dissociating ultraviolet (UV) radiation field, and a simple time-independent analytic approximation. We show that these two methods produce excellent agreement at all metallicities $\ga 10^{-2}$ of the Milky Way value across a very wide range of UV fields. At lower metallicities the agreement is worse, likely because time-dependent effects become important; however, there are no observational calibrations of molecular gas content at such low metallicities, so it is unclear if either method is accurate. The comparison suggests that, in many but not all applications, the analytic approximation provides a viable and nearly cost-free alternative to full time-dependent chemistry and radiative transfer.
\end{abstract}


\section{Introduction}
\label{sec:intro}

Resolved observations of nearby galaxies in the last decade have unequivocally established that the star formation rate in a galaxy correlates much more strongly with its molecular content than with its total gas content \citep{wong02a, kennicutt07a, bigiel08a, leroy08a, blanc09a}. Driven by these observations, analytic models \citep{blitz06b, krumholz09b}, semi-analytic models \citep{obreschkow09b, obreschkow09a, fu10a, dutton10a}, and simulations \citep{pelupessy06a, robertson08b, pelupessy09a, ng:gtk09, ng:gk10a, ng:gk10b} have all begun to incorporate models for the \hi\ to H$_2$ transition into their recipes for star formation. The inclusion of molecule formation has allowed these models to solve a number of problems that appeared in earlier models that omitted molecules. For example in the high redshift universe, absorption line surveys indicate that the star formation rate in damped Lyman $\alpha$ systems must be significantly below what one would expect given their \hi\ column densities if the star formation rate were correlated with total gas surface density as it is in nearby galaxies \citep{wolfe06a, wild07a}. This can be explained by the fact that these systems are usually quite metal poor, which causes them to be systematically deficient in molecular hydrogen and thus in star formation compared to local metal-rich galaxies of similar total gas column density \citep{krumholz09e,ng:gk10b}. Similarly, in the local universe, models incorporating molecule formation are able to explain why some galaxies that undergo \hi\ stripping in a cluster environment have their star formation quenched, while others do not. Those that lose their outer gas disks but retain high column density, molecule-rich centers continue to form stars, while those that lose their inner gas disks cease to form molecules and stars \citep{fumagalli09a}.

While many of these authors have compared their models or simulations to observed molecular fractions in nearby galaxies, these observations cover only a limited dynamic range in metallicity, radiation environment, and other quantities that may be relevant to the \hi\ to H$_2$ transition. We cannot yet test the models in more extreme environments with observations, but we can check for consistency between the models. Such checks provide at least a minimum level of confidence in extrapolating the model predictions beyond the range of environments in which they have been calibrated. Cross-comparisons are particularly important given the range of complexity of the models, which go from steady-state analytic approximation formulae to sophisticated time-dependent radiation plus chemistry modules. Only the former are suitable for implementation in semi-analytic models, and thus such a test is necessary to ensure that semi-analytic models that rely on them are able to faithfully reproduce the results of numerical simulations.

Our goal in this paper is to provide such an inter-model comparison. We use a suite of numerical models by \citet[hereafter GK10; see \S~\ref{sec:numerical} for a detailed description]{ng:gk10b}, and we compare them to one of the most commonly-used analytic approximations \citep[see \S\ref{sec:analytic} for a detailed description]{krumholz08c, krumholz09a, mckee10a} for the \hi\ to H$_2$ transition. Both of the analytic and numerical models have been tested against observations of local galaxies, and have been found to provide a good description of the molecular content in them.\footnote{In contrast, we do not compare to the H$_2$-midplane pressure correlation model of \citet{blitz06b}, because recent observations indicate that it breaks down on the sub-kpc scales resolved in the simulations \citep{fumagalli10a}.} This comparison will allow us to establish under what circumstances the numerical and analytic models agree and thus can be applied with reasonable confidence, and to gain physical insight into the H$_2$ formation process by examining cases where they differ. An additional practical benefit is that, in cases where the analytic models are able to do a reasonable job reproducing the numerical ones, they may provide an approximation that can be used in simulations at greatly reduced computational cost.

The structure of the remainder of this paper is as follows. In \S~\ref{sec:methods} we summarize the numerical and analytic models for the \hi\ to H$_2$ transition. In \S~\ref{sec:fixedrad} we compare the models in idealized simulations in which the radiation field and metal content are held fixed, while in \S~\ref{sec:scrad} we consider simulations in which the radiation field and metal content are determined in a fully self-consistent manner. Finally in \S~\ref{sec:summary} we discuss and summarize our conclusions.

\section{Summary of the Methods}
\label{sec:methods}

Here we summarize how we estimate the H$_2$ mass fraction in every computational cell of our test simulations. Section \ref{sec:numerical} describes the time-dependent numerical chemistry method, and Section \ref{sec:analytic} describes the analytic model and how we apply it to the simulations.

\subsection{The Numerical Model}
\label{sec:numerical}

Simulations used in this paper are fully described in GK10. Here we only remind a reader that a set of ``fixed ISM'' simulations was performed with the Adaptive Refinement Tree (ART) code \citep{sims:k99,sims:kkh02,sims:rzk08}. Each simulation was started from a $z=4$ output of a fully self-consistent cosmological simulation of a Milky Way progenitor galaxy. However, in each fixed ISM simulation the dust-to-gas ratio and the interstellar radiation field were fixed to pre-defined, constant in space and time values, effectively turning a cosmological simulation into a highly realistic simulation of an isolated galaxy with a live dark matter halo and natural accretion history.

Each fixed ISM simulation was continued for 600 Myr, long enough to fully establish a new ISM structure and eliminate the memory of the initial conditions. The dark matter mass resolution in the simulation is $1.3\times10^6\Msun$ and the spatial resolution of the highest resolved region of the simulation (that includes the whole of the simulated galaxy disk) is 65 pc. That translates into the mass resolution in the gas of between $\sim 10^3\Msun$ and $\sim 10^6\Msun$, depending on the value of the gas density.

All fixed ISM simulations include gas dynamics, star formation, and the time-dependent and spatially-inhomogeneous 3D radiative transfer of ultraviolet and ionizing radiation from individual stellar particles that formed during the course of each simulation. The simulations incorporate a non-equilibrium chemical network of hydrogen and helium and non-equilibrium cooling and heating rates, which make use of the local abundance of atomic, molecular, and ionic species and UV intensity.  This network includes formation of molecular hydrogen both in the primordial phase and on dust grains. The abundances of the relevant atomic and molecular species are therefore followed self-consistently during the course of the simulation. The heating and cooling terms in the equation for the internal energy include all of the terms normally included in the simulations of first stars and in the ISM models, including cooling on metals. A complete set of chemical reactions and the values for the adopted reaction and cooling/heating rates are presented in the appendix of GK10.

In addition to the simulations with the fixed ISM properties, we also use a fully self-consistent cosmological simulation of several galaxies in a cosmological volume with $25h^{-1}$ Mpc on a side. These simulations will be fully described elsewhere (M.\ Zemp et al, in preparation), but in physical modeling they are identical to the simulation of \citet{ng:gk10a}; the only difference from the early simulation is a larger box size and slightly different values of cosmological parameters, based on the WMAP5 cosmology \citep{cosmo:dkns08}. The spatial and mass resolution of the cosmological simulation is matched exactly to the spatial and mass resolution of the simulation with the fixed ISM. Due to computational expense, the fully self-consistent cosmological simulation was not continued beyond $z=3$. For comparison with the analytical model, we use several of the model galaxies from that simulation that differ widely in their ISM properties. In model galaxies from a full cosmological simulation the properties of the ISM (dust-to-gas ratio, gas metallicity, the interstellar radiaton field, the gas density distribution, etc.) vary inside the galaxy in a manner that is controlled by their prior cosmic history and the present cosmological environment.

Finally, we call the reader's attention to two details of the microphysical model of H$_2$ formation in the simulations that will be important for our comparison. First, since the H$_2$ destruction rate depends on the amount of shielding, one must estimate a column density for each cell. In the simulations one does this using the Sobolev approximation: in each cell one computes the local density scale height $h = \rho/|\nabla \rho|$, and then takes the column density to be $\Sigma=\rho h$. Comparison of this approximation with ray-tracing done in post-process shows that it is generally quite accurate. Second, one must increase the rate coefficient for H$_2$ formation by a clumping factor to account for density inhomogeneities that are unresolved on the computational grid. Since the H$_2$ formation rate varies as the square of density, these inhomogeneities increase the overall rate. The simulations use a clumping factor of 30, which is chosen to give a good match between the simulations and FUSE observations of the Milky Way and Magellanic clouds.

\subsection{The Analytic Model}
\label{sec:analytic}

The analytic model to which we compare the simulations is based on the work of \citet{krumholz08c}, \citet{krumholz09a}, and \citet{mckee10a}, hereafter known as the KMT model. KMT consider an idealized spherical cloud immersed in a uniform, isotropic Lyman-Werner band radiation field. They then solve the coupled problems of radiative transfer and \htwo\ formation-dissociation balance under the assumption that the cloud is in steady state. They do not consider gas-phase reactions (though see the Appendix of \citealt{mckee10a} for a discussion of how these might be included). The solution is
\begin{equation}
\label{eq:fh2anlayt}
\fhtwo \simeq 1 - \left(\frac{3}{4}\right) \frac{s}{1+0.25s},
\end{equation}
where $\fhtwo$ is the \htwo\ mass fraction in the cloud,
\begin{eqnarray}
s & = & \frac{\ln(1+0.6\chi + 0.01\chi^2)}{0.6 \tau_c} \\
\chi & = & 71 \left(\frac{\sigma_{d,-21}}{\calr_{-16.5}}\right) \frac{G_0'}{\nhzero},
\label{eq:chi1}
\end{eqnarray}
$\tau_c$ is the dust optical depth of the cloud, $\sigma_{d,-21}$ is the dust cross-section per H nucleus to 1000 \AA\ radiation, normalized to a value of $10^{-21}$ cm$^{-2}$, $\calr_{-16.5}$ is the rate coefficient for H$_2$ formation on dust grains, normalized to the Milky Way value of $10^{-16.5}$ cm$^3$ s$^{-1}$ \citep{wolfire08a}, $G_0'$ is the ambient UV radiation field intensity, normalized to the \citet{draine78} value for the Milky Way, and $\nhzero$ is the volume density of H nuclei in units of cm$^{-3}$. Note that $\sigma_d$ and $\calr$ are both proportional to the dust to gas ratio, so their ratio is independent of the metallicity.

In order to apply this analytic model to the simulations, we must have a way of computing the two dimensionless numbers $\chi$ and $\tau_c$ on which $\fhtwo$ depends. Our estimate of $\tau_c$ is straightforward. Since the resolution of the simulations is well-matched to individual molecular clouds, we estimate the local column density $\Sigma$ from the Sobolev approximation exactly as the simulations do (see Section \ref{sec:numerical}). The dust optical depth is simply $\tau_c = \Sigma \sigma_d/\mu_{\rm H}$, where $\sigma_d$ is the dust cross section per H nucleus and $\mu_{\rm H}=2.3\times 10^{-24}$ g is the mean mass per H nucleus. We take the cross section to have a value $\sigma_{d,-21} = Z'$, where $Z'$ is the metallicity normalized to the Milky Way value. For $Z'$ we use the local metallicity in the cell for which we are computing $\tau_c$.

The scaled radiation field $\chi$ requires a bit more care. In Section \ref{sec:fixedrad} we discuss simulations where the UV field is fixed to a specified value rather than determined self-consistently. In that case we compute $\chi$ directly from equation (\ref{eq:chi1}) using the UV field, gas volume density, dust cross section, and rate coefficient adopted in the simulation. The dust cross section is again $\sigma_{d,-21} = Z'$, while the rate coefficient is $\calr_{-16.5} = 30 Z'$. This value of $\calr$ includes the factor of 30 enhancement due to unresolved clumping adopted in the simulations (see Section \ref{sec:numerical}).

In contrast, in Section \ref{sec:scrad} we compare to simulations where the UV field is determined self-consistently and varies from point to point. In this case we can proceed in two ways. First, we can use the UV field computed in the simulations in every cell, in analogy to the first case. Second, and more interestingly, we can omit this information. KMT show that, if the interstellar medium is in a self-consistently determined two phase equilibrium, the ratio $G_0'/\nhzero$ should take on a characteristic value such that
\begin{equation}
\label{eq:chi2}
\chi \approx 3.1 \left(\frac{1+3.1 Z'^{0.365}}{4.1}\right),
\end{equation}
where $Z'$ is the metallicity relative to solar units. This is not likely to hold cell-by-cell at every time step, but KMT argue that it should hold on average. The advantage of this approach is that using this value of $\chi$ can we can compute the H$_2$ fraction using only the local metallicity and the column density as determined by the Sobolev approximation. Thus this method can be used in simulations or semi-analytic models that do not explicitly include either chemistry or radiative transfer. Since the radiative transfer and chemistry are often the most computationally costly parts of a simulation, it would be extremely useful to be able to roughly approximate the results of a full chemistry and radiation simulation with such a simple analytic model.

\section{Simulations with Fixed ISM Properties}
\label{sec:fixedrad}

\begin{figure*}
\plotone{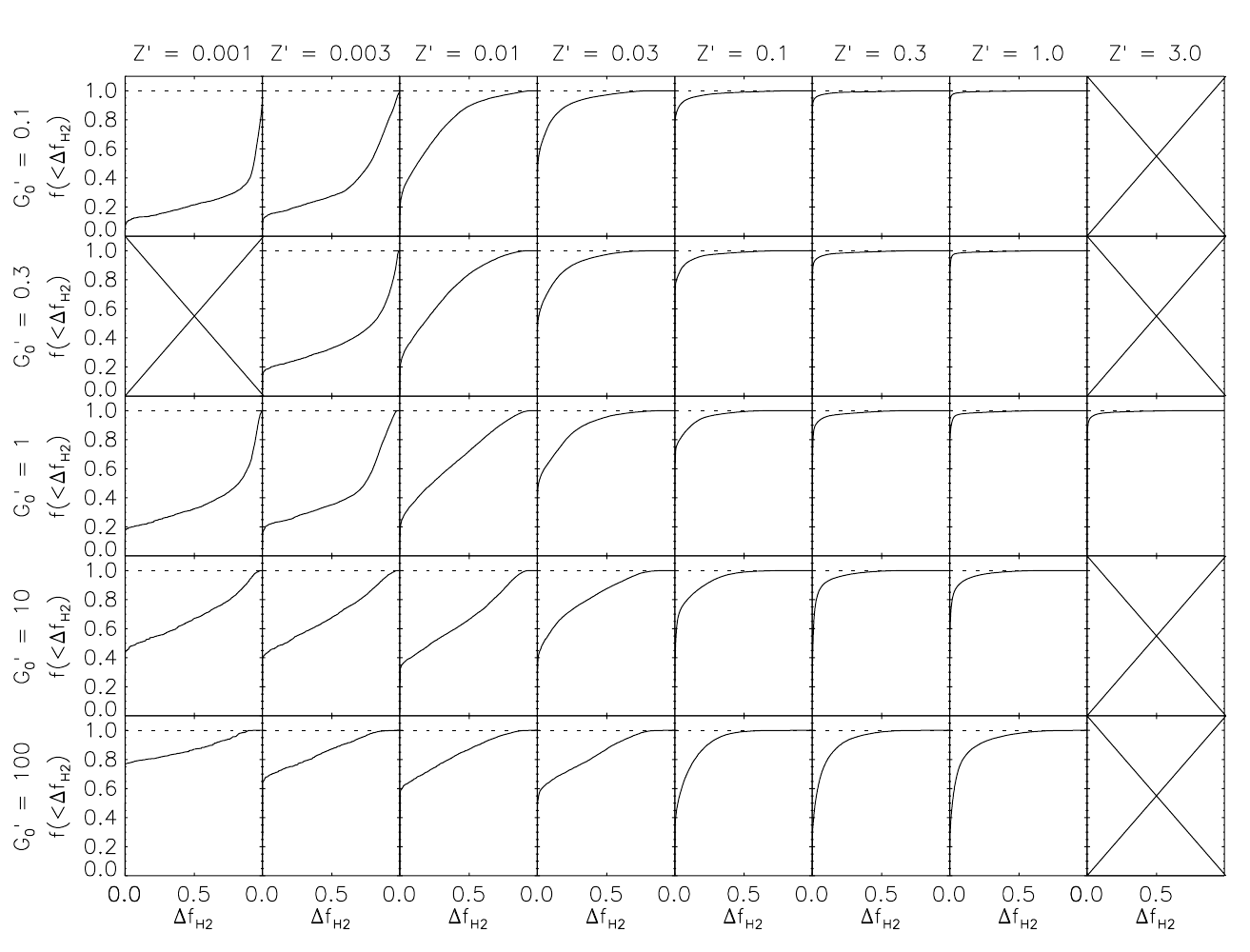}
\caption{\label{fig:h2compgrid1}
Cumulative fraction of the total simulation mass $f(<\dfhtwo)$ for which the difference between the simulation and analytically-predicted values of $\fhtwo$ is less than $\dfhtwo$. Columns show simulations of fixed metallicity relative to Milky Way, $Z'$, as indicated at the top of each column, and rows show simulations of fixed UV radiation field relative to Milky Way, $G_0'$, as indicated at the left of each row. Dashed horizontal lines show a value of $1.0$, indicating perfect agreement. An X in a box indicates that no simulation for that value of $Z'$ and $G_0'$ is available.
}
\end{figure*}

In order to gain a better understanding of how the model and simulations compare, we begin by considering a set of simulations in which the radiation field and metallicity are specified by hand to have a particular value, rather than being determined self-consistently. The grid of metallicities runs from $Z'=10^{-3} - 3$, and the grid of UV fields runs from $G_0' = 0.1 - 100$. From each simulation we extract all computational cells that are located in the disk of the simulated galaxy. For each computational cell we record its total hydrogen density, temperature, metallicity, gas-to-dust ratio, local value of the interstellar radiation field, and abundances of ionized, atomic, and molecular hydrogen. 

To quantify the level of agreement between the simulations and the analytic model, in every computational cell we use the analytic model described in Section \ref{sec:analytic} to compute a predicted H$_2$ fraction $\fhtwoanalyt$. We compute the absolute difference between this and the value obtained in the simulation $\fhtwosim$,
\begin{equation}
\dfhtwo = |\fhtwoanalyt - \fhtwosim|,
\end{equation}
because the star formation rate depends on the absolute value of the molecular abundance, not on relative one. For example, regions with $\fhtwo \ll 1$ may show a large relative error between the two models, but those regions would not contribute to star formation rate in a galaxy or on a particular spatial scale.

Figure \ref{fig:h2compgrid1} shows $f(<\dfhtwo)$, the fraction of the mass in the extracted galaxy disk cells (not in the entire cosmological simulation) for which the difference is less than $\dfhtwo$. In this plot perfect agreement would appear as a function of constant value unity, while complete disagreement would be a function of constant value zero. The Figure indicates that the agreement is extremely good for all UV field strengths at metallicities $Z'\ga 10^{-2}$. In all of these cases the disagreement is essentially zero for at least half of the mass, and is less than 25\% for $\ga 75\%$ of the mass. The Milky Way-like case $Z'=1$ and $G_0'=1$ shows near-perfect agreement between the simulations and the analytical model. In contrast, the agreement is quite poor for the lower metallicity cases, particularly for the combination of low metallicity and weak radiation field.

In order to understand why some simulations agree and others do not, we construct bins of $\fhtwosim$ running from $0-0.2$, $0.2-0.4$, $0.4-0.6$, $0.6-0.8$, and $0.8-1.0$, and similarly for $\fhtwoanalyt$. We then compute what fraction of the total mass in the simulation falls into a given bin of $\fhtwosim$ versus $\fhtwoanalyt$. If agreement were perfect all the mass would fall into the same bin on either axis, i.e.\ all the mass with $\fhtwosim$ in the range $0-0.2$ would also have $\fhtwoanalyt$ in this range, and so forth. 

\begin{figure*}
\plotone{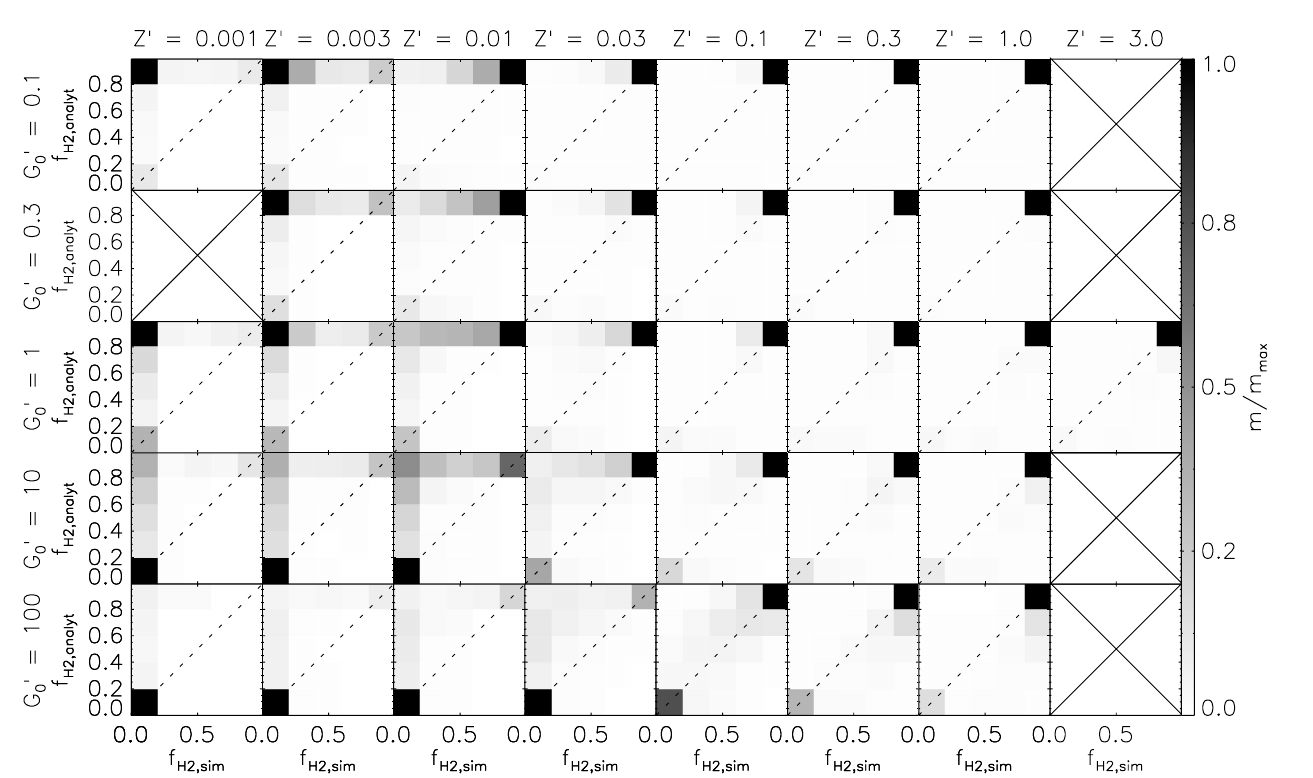}
\caption{\label{fig:h2compgrid2}
In each panel the colors pixels indicate the fraction of the mass of that simulation in the given bin of $\fhtwosim$ versus $\fhtwoanalyt$, divided by the fraction of mass in the most massive bin for that simulation, so that the bin in each panel that contains the most mass is always black. See text for details. Perfect agreement would correspond to all pixels lying along the dashed one-to-one lines through the center of each panel. Simulations are labeled with values of $Z'$ and $G_0'$ as in Figure \ref{fig:h2compgrid1}.
}
\end{figure*}

\begin{figure*}
\plotone{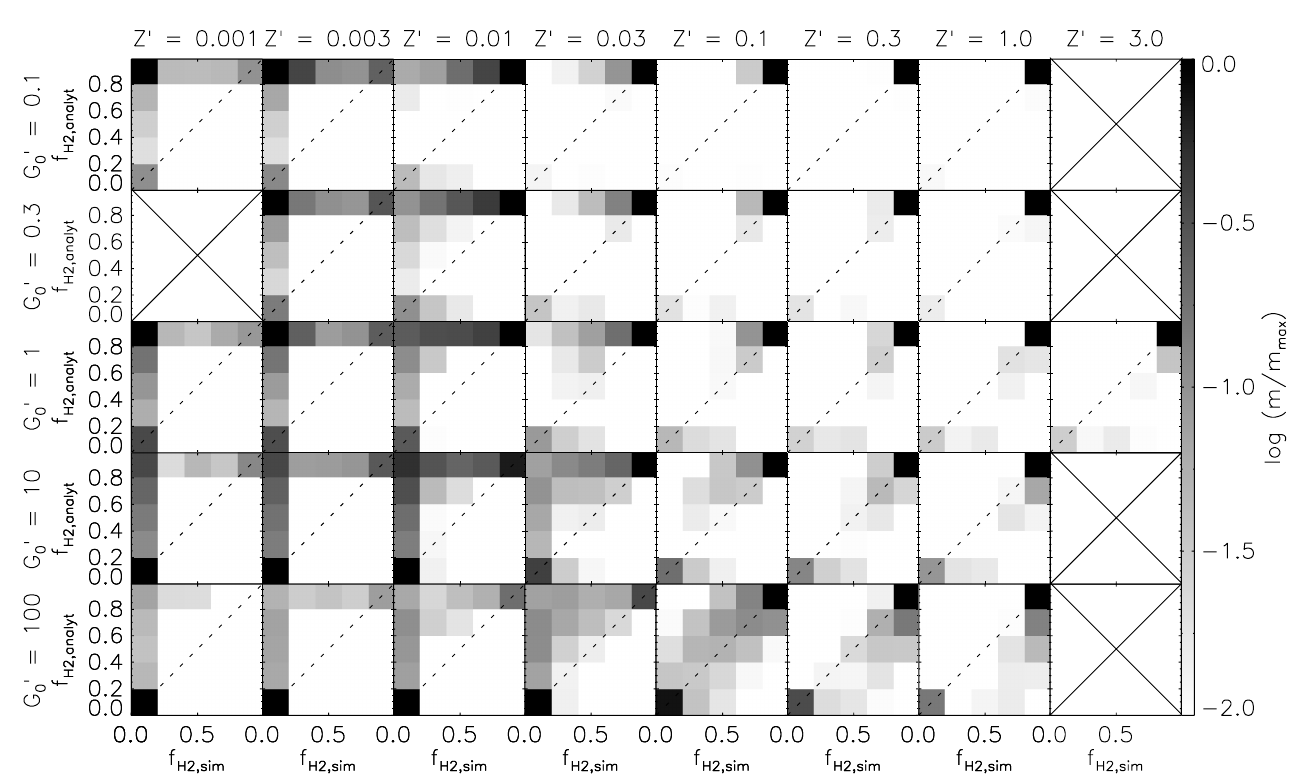}
\caption{\label{fig:h2compgrid3}
Same as Figure \ref{fig:h2compgrid2}, but using a logarithmic rather than a linear scale to emphasize areas of disagreement.
}
\end{figure*}

We plot the actual distribution of mass in bins in Figures \ref{fig:h2compgrid2} and \ref{fig:h2compgrid3}. The former figure shows a linear color scale, indicating where the bulk of the mass lies, while the latter shows a logarithmic scale, intended to emphasize areas of disagreement even if they contain relatively little mass. The figures reveal some interesting systematic differences between the simulations where the analytic model agrees well with the numerical results, at $Z'>10^{-2}$, and the ones where it does not, at $Z'<10^{-2}$. First, in the higher metallicity cases the majority of the mass in both the simulations and the analytic models tends to be mostly molecular or mostly atomic; bins of intermediate molecular fraction are sparsely populated. In contrast, at lower metallicity a significant fraction of the mass is in the intermediate molecular fraction regime, and here the analytic model tends to systematically overestimate the molecular fraction compared to the simulation.

\begin{figure}
\includegraphics[width=\hsize]{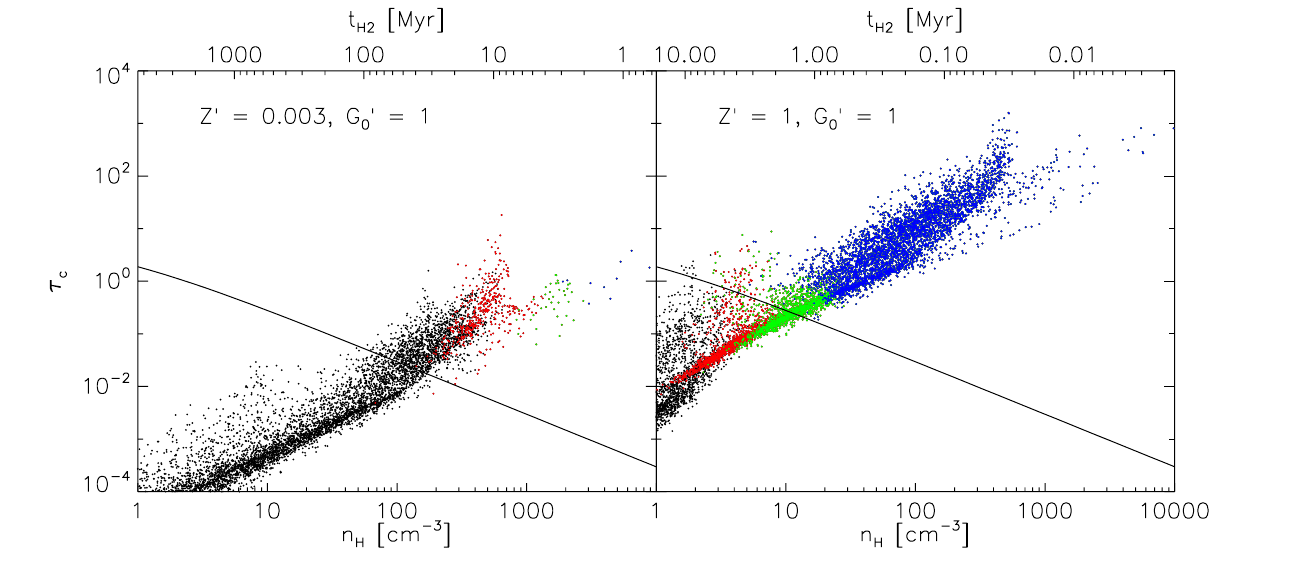}
\caption{\label{fig:h2scatter}
Each point shows the value of $\nh$ and $\tau_c$ in a computational cell in the simulation with the indicated values of $Z'$ and $G_0'$. (To minimize file size we show only a subset of cells.) Points are color-coded by molecular fraction in the simulation, with black indicating $\fhtwosim<0.1$, red indicating $0.1<\fhtwosim<0.5$, green indicating $0.5<\fhtwosim<0.9$, and blue indicating $0.9<\fhtwosim$. In contrast, the solid line separates points with $\fhtwoanalyt<0.5$ (below the line) from those with $\fhtwoanalyt>0.5$ (above the line).
}
\end{figure}

Although the cause of the discrepancy is not entirely clear, we regard the most likely explanation to be that the steady-state assumption in the model is breaking down at metallicities $Z'<10^{-2}$. Consider a computational cell of volume density $\nh$ that is predominantly atomic. If the dissociation rate is negligible, the time required for that cell to convert to molecular is the ratio of the \hi\ volume density to the volumetric rate of conversion of \hi\ to \htwo\
\begin{equation}
t_{\rm H_2} = \frac{\nh}{2\nh^2 \calr} = 17 \nhzero^{-1} Z'^{-1} \mbox{ Myr},
\end{equation}
where we adopt the rate coefficient $\calr_{-16.5} = 30 Z'$ used in the simulations.

Figure \ref{fig:h2scatter} illustrates the implications of this timescale by comparing one simulation where the analytic model agrees very well with the simulation H$_2$ fraction ($Z'=1$, $G_0'=1$) to another where they agree poorly ($Z'=0.003$, $G_0'=1$). In the Milky Way metallicity case the locus where the analytic model predicts 50\% H$_2$ fraction intersects the region of parameter space occupied by the simulation points near $\nh\approx 10$ cm$^{-3}$, corresponding to an H$_2$ formation time $t_{\rm H_2} \sim 2$ Myr at $Z' = 1$. This is short compared to any galactic timescale, and thus the instantaneous equilibration approximation is an excellent one. On the other hand, in the low metallicity simulation the locus of intersection is at $\nh\approx 100$ cm$^{-3}$, corresponding to $t_{\rm H_2}\sim 100$ Myr at $Z' = 0.003$. In this case a parcel of gas may be compressed in a spiral arm to the point where one might expect it to become H$_2$-dominated were it able to reach steady state, but before it has a chance to do so it leaves the spiral arm and its density and column density fall. The analytic model identifies such regions as predominantly molecular because it ignores the time dependence.

Before concluding that the simulations are correct and the model erroneous in this case, however, we must add a final caution. The timescale we compute is inversely proportional to the subgrid clumping factor assumed in the simulations. While this number has been estimated for metallicities down to $Z'\approx 0.2$ in the Small Magellanic Cloud, we have no direct knowledge of its value in lower metallicity galaxies. Nonetheless, we tentatively conclude that the time-dependence of the H$_2$ fraction is not a major effect at metallicities $Z' \ga 10^{-2}$, so a time-independent model is adequate in this regime.

\section{Simulations with Self-Consistently Determined Radiation Fields}
\label{sec:scrad}

\begin{deluxetable*}{ccccccccc}
\tablecaption{
\label{tab:galaxies}
Summary of Galaxy Properties
}
\tablehead{
\colhead{Galaxy} &
\colhead{Halo mass ($\msun$)} &
\colhead{Stellar mass ($\msun$)} &
\colhead{Gas mass ($\msun$)} &
\colhead{$\langle f_{\rm HI}\rangle$} &
\colhead{$\langle f_{\rm H_2}\rangle$} &
\colhead{$\langle Z' \rangle$} &
\colhead{$\langle G_0' \rangle_M$} &
\colhead{$\langle G_0' \rangle_V$}
}
\startdata
1 & $4.8\times10^{11}$ & $4.4\times10^{10}$ & $4.6\times10^{10}$ & 0.09 & 0.02 & 0.5 & 1{,}100 & 92 \\
2 & $2.3\times10^{10}$ & $5.3\times10^{7}$ & $5.9\times10^{9}$ & 0.37 & 0.09 & $0.01$ & 160 & 17 \\
3 & $2.9\times10^{10}$ & $1.0\times10^{9}$ & $4.5\times10^{9}$ & 0.26 & 0.15 & $0.18$ & 39 & 7.4 \\
\enddata
\tablecomments{Col.\ 5-8: Mass-weighted average H~\textsc{i} fraction and H$_2$ fraction within the virial radius of a galaxy, relative metallicity of neutral gas, and relative radiation intensity in the neutral gas (mass and volume weighted). Note that $\langle f_{\rm HI}\rangle + \langle f_{\rm H_2}\rangle < 1$ because the gas mass budget includes ionized hydrogen as well.
}
\end{deluxetable*}

We now proceed to compare the KMT model to simulations that include an explicit, self-consistent calculation of the local radiation field, and use this local radiation field to compute the H$_2$ dissociation rate. For this comparison we select three galaxies from the simulations described in \S~\ref{sec:numerical} at redshift $z= 3$, chosen to represent a wide range of galactic types and environments. Table \ref{tab:galaxies} lists some of the basic properties of the three galaxies. Galaxy 1 is a typical $z=3$ star-forming galaxy, with a high radiation field, relatively high metallicity, and generally high gas densities; by $z=0$ it is expected to evolve into a Milky Way type galaxy. In contrast, galaxy 2 has just started its first starburst. As a result, it has a fairly strong radiation field, but it has not yet produced many metals and thus has a very low metallicity. Finally, galaxy 3 is a dwarf, SMC-like galaxy that has been forming stars at a low rate, and remains mostly gaseous. As a result is has the lowest radiation field of any of the three, but a fairly high metallicity.

\begin{figure}
\plotone{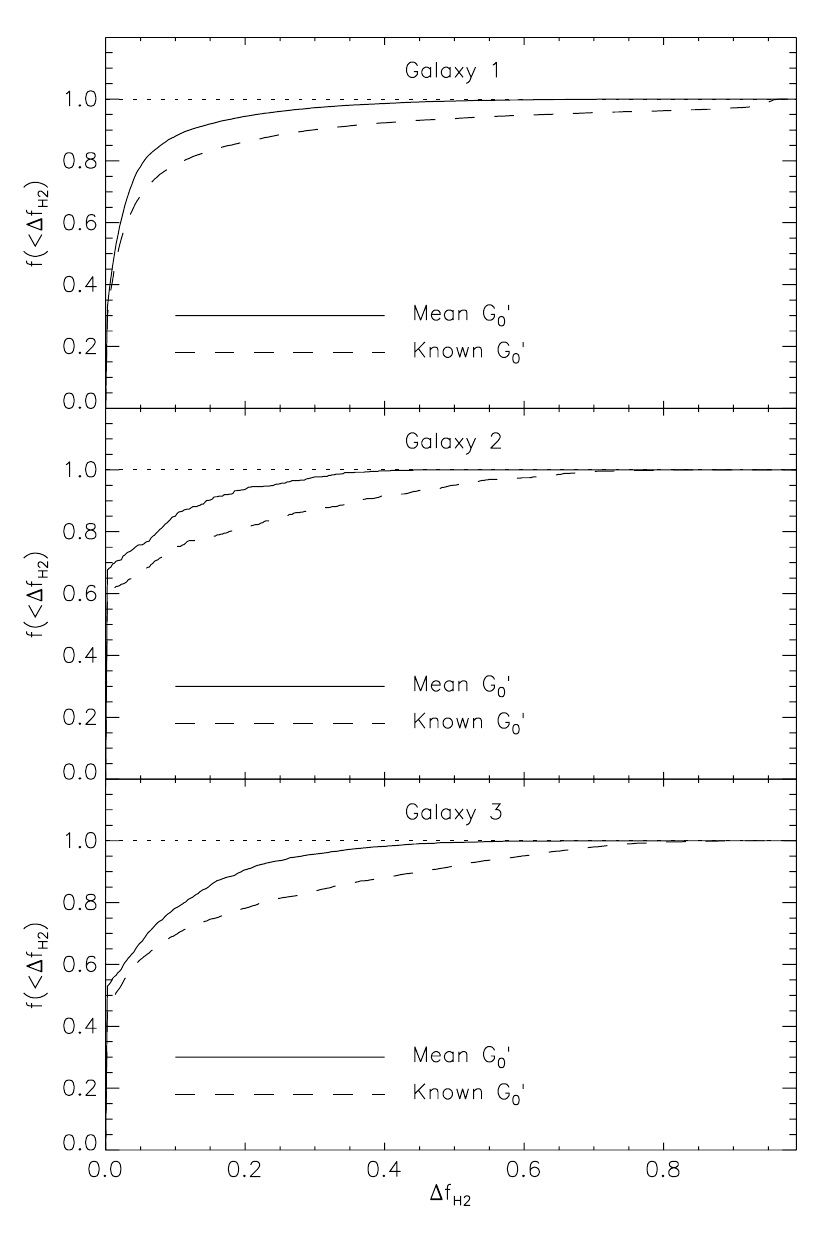}
\caption{\label{fig:sccomp1}
Same as Figure \ref{fig:h2compgrid1}, but for the three sample galaxies with self-consistently computed radiation fields and H$_2$ fractions. Dashed lines show the result when we compute $f_{\rm H_2}$ in the KMT model using the local radiation field in each cell, while solid lines show the result when we compute $f_{\rm H_2}$ in the KMT model without using any information about the local radiation field, and we instead use the mean radiation field given by equation (\ref{eq:chi2}).
}
\end{figure}

Figure \ref{fig:sccomp1} shows, for each of the three sample galaxies, the cumulative fraction of the simulation mass for which the difference between the simulation and the KMT model is less than $\Delta f_{\rm H_2}$. The dashed line corresponds to the case where we compute $\Delta f_{\rm H_2}$ using our knowledge of the local radiation field, while the solid line corresponds to adopting the mean radiation field given by equation (\ref{eq:chi2}). The latter is the model that can be applied in a simulation that does not include a radiation and chemistry module.

The first thing to notice is that, for all three galaxies, the KMT model using the mean radiation field (solid lines) agrees very well with the simulation result. In all cases the difference in H$_2$ fraction between the model and simulation is less than 10\% for $\sim 80\%$ of the mass, and is less than 20\% for $\sim 90\%$ of the mass. This indicates that we are able to reproduce the simulation results very well, across a wide range of galaxies, using a model that requires no knowledge of the local radiation field. This is highly encouraging for numerical simulations and analytic and semi-analytic models, since it suggests that the results of a full radiative transfer and chemistry solve can be approximated reasonably well at far less computational cost.

\begin{figure}
\plotone{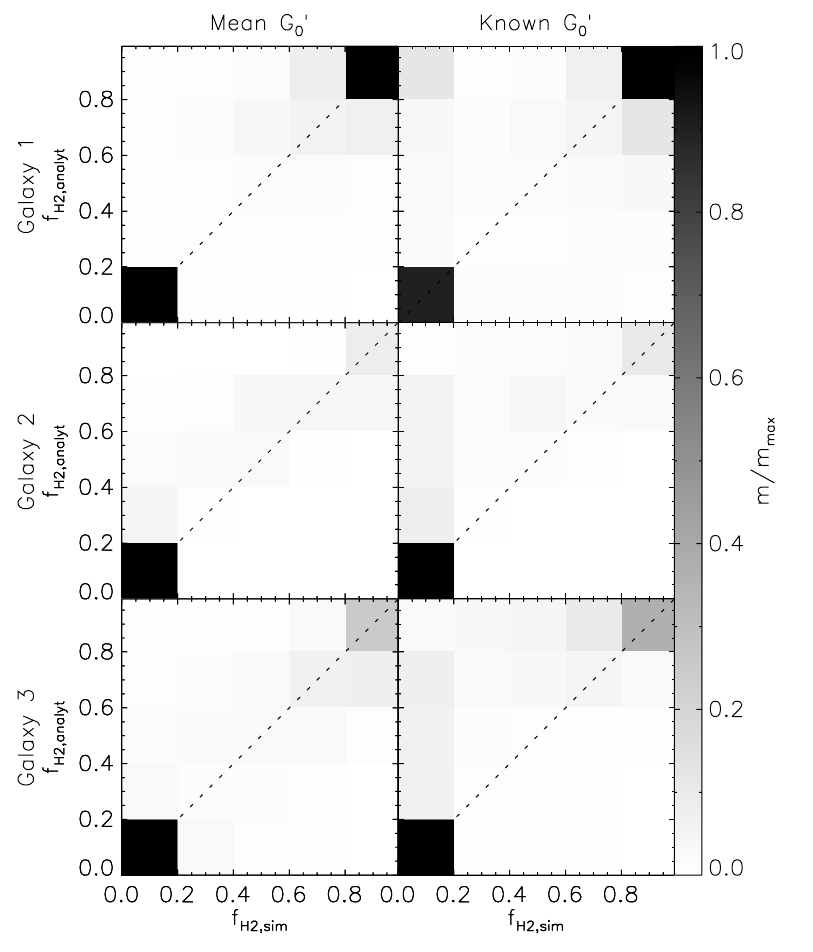}
\caption{\label{fig:sccomp2}
Same as Figure \ref{fig:h2compgrid2}, but for the three sample galaxies with self-consistently computed radiation fields and H$_2$ fractions. The left column shows the results using the mean radiation field given by equation (\ref{eq:chi2}), while the right column gives the results using the local radiation field in each cell computed in the simulation.
}
\end{figure}

\begin{figure}
\plotone{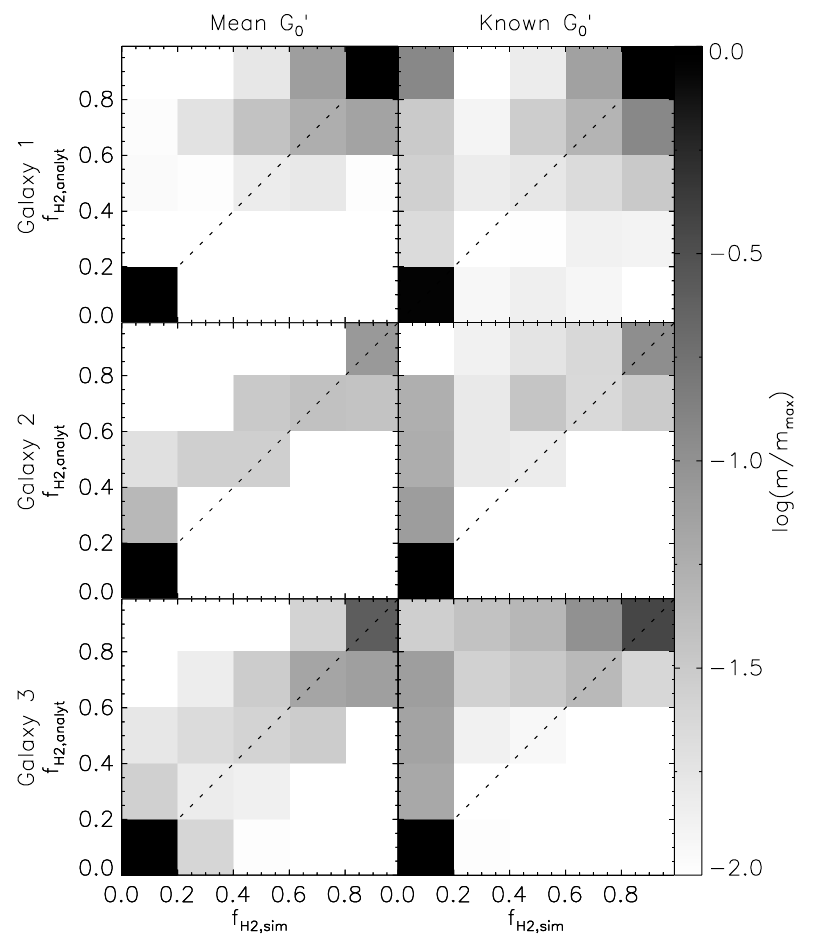}
\caption{\label{fig:sccomp3}
Same as Figure \ref{fig:sccomp2}, but using a logarithmic rather than a linear scale to emphasize areas of disagreement.
}
\end{figure}

Surprisingly, the agreement between the simulations and the KMT model actually worsens if we do use our knowledge of the local radiation field. To help understand why this is, in Figures \ref{fig:sccomp2} and \ref{fig:sccomp3} we show plots analogous to those in Figures \ref{fig:h2compgrid2} and \ref{fig:h2compgrid3} for the self-consistent case. As in Figures \ref{fig:h2compgrid2} and \ref{fig:h2compgrid3}, we take each simulation and divide the mass into bins of $f_{\rm H_2,sim}$ and $f_{\rm H_2,analyt}$, and plot how much mass falls into each bin; perfect agreement would then appear as all colored pixels falling along the diagonal $1-1$ line. The plots show that, in the case where we adopt the mean radiation field of equation (\ref{eq:chi2}), most of the mass is indeed in pixels along the $1-1$ line, and mass that is off this line is scattered fairly symmetrically about it. In contrast, if we use the instantaneous, known radiation field value in each cell, we find that the KMT model systematically overestimates the molecular fraction relative to the simulation. This effect is smallest in galaxy 1, which has the highest metallicity, and is larger in galaxies 2 and 3, which have lower metallicity.

This is very likely another version of the timescale effect we identified in Section \ref{sec:fixedrad}. The timescale to reach equilibrium in the lower metallicity simulations is not much smaller than the timescale over which radiation fields and densities fluctuate, and thus the relevant value of $\chi$ is a time-averaged rather than an instantaneous one. The instantaneous value of $G_0'$ is usually smaller than its time-average (because there are few regions of intense radiation and many regions of little radiation), and so using this instantaneous value leads to an overestimate of the H$_2$ fraction in those cases where the metallicity is low enough to make the equilibration time long. It is interesting, however, that this effect seems to be mitigated if we simply adopt the mean characteristic radiation field expected for two-phase equilibrium. In this case we obtain good agreement even with for the lowest metallicity self-consistent galaxy, which has $\langle Z'\rangle = 0.01$.

\section{Summary}
\label{sec:summary}

In this work we compare two approaches to the problem of determining where the ISM in galaxies transitions from \hi\ to H$_2$. One is a highly-accurate time-dependent chemistry and radiative transfer method implemented in a large-scale cosmological simulation code, while the other is a simple analytic approximation suitable for implementation in semi-analytic models or simulation codes that to not include chemistry or radiative transfer. Our comparison shows that the two models agree extremely well whenever the metallicity is of order 1\% of the Solar value or more. We find this agreement in both experimental galaxies with fixed metallicity and radiation field, and in galaxies whose metallicity and radiation field are computed in a fully self-consistent manner. The two models diverge at even lower metallicities, and we interpret this as being a result of a failure of the assumption of chemical equilibrium in the analytic model. At such low metallicities, the time required to convert a given fluid element from \hi\ to H$_2$ is comparable to a galactic rotation period even in the absence of dissociating radiation, and the equilibrium model overestimates the H$_2$ fraction in this case. 

Nonetheless, the excellent agreement we find at metallicities of 1\% Solar or more suggests that the analytic model has its uses. In simulations, it is capable of reproducing the results of the much more complex and accurate time-dependent chemistry and radiation module at essentially zero cost in terms of both computational time and programming effort. The analytic model also allows analytic or semi-analytic models to incorporate the effects of the \hi\ to H$_2$ transition in a way that faithfully reproduces the results of simulations. For these reasons it is a valuable tool for cosmology.

\acknowledgements
Support for this work was provided by: an Alfred P. Sloan Fellowship (MRK); NSF grants AST-0807739 (MRK), AST-0908063 (NYG), and CAREER-0955300 (MRK); NASA through Astrophysics Theory and Fundamental Physics grants NNX09AK31G (MRK) and NNX-09AJ54G (NYG), and a Spitzer Space Telescope Theoretical Research Program grant (MRK). We also thank the Aspen Center for Physics, where much of the work for this paper was performed.

\bibliographystyle{apj}
\bibliography{refs,ng-bibs/self,ng-bibs/sims,ng-bibs/cosmo}

\end{document}